\documentclass[aps,prd,amsmath,11pt,superscriptaddress,floatfix,nofootinbib]{revtex4-1}
\usepackage{epsfig,amssymb,amsfonts,amsmath,mathtools,bm,color,xcolor,graphicx,braket,adjustbox,esint,upgreek}
\usepackage{multirow}
\usepackage[utf8]{inputenc}
\usepackage{setspace}
\usepackage{dcolumn,kantlipsum}
\usepackage{rotating}

\usepackage{resizegather}
\allowdisplaybreaks

\usepackage{natbib}
\usepackage{graphicx}

\newcommand{\be}{\begin{equation}}
\newcommand{\ee}{\end{equation}}
\newcommand{\bea}{\begin{eqnarray}}
\newcommand{\eea}{\end{eqnarray}}
\newcommand{\beas}{\begin{eqnarray*}}
\newcommand{\eeas}{\end{eqnarray*}}

\def\vec#1{\boldsymbol{#1}}

\begin{document}

\title{The nature of charged charmonium-like states $Z_c(3900)$ and its strange partner $Z_{cs}(3982)$}

\author{Meng-Chuan Du$^{2,3}$\footnote{{\it E-mail address:} dumc@ihep.ac.cn}, Qian Wang$^{1,2}$\footnote{{\it E-mail address:} qianwang@m.scnu.edu.cn; Correspondence author},
and Qiang Zhao$^{2,3}$\footnote{{\it E-mail address:} zhaoq@ihep.ac.cn; Correspondence author}
}

\affiliation{ Guangdong Provincial Key Laboratory of Nuclear Science, \\
Institute of Quantum Matter, South China Normal University, Guangzhou 510006, China}
\affiliation{Guangdong-Hong Kong Joint Laboratory of Quantum Matter, Southern Nuclear Science Computing Center, South China Normal University, Guangzhou 510006, China}
\affiliation{ Institute of High Energy Physics, Chinese Academy of Sciences, Beijing 100049, China}
\affiliation{ University of Chinese Academy of
Sciences, Beijing 100049, China}

\begin{abstract}
The observation of $Z_{cs}(3985)$ by BESIII in the $D_s^{*-}D^0$ and  $D_s^{-}D^{*0}$ channel adds important dynamic information for a coherent understanding of both $Z_c(3900)$ and $Z_c(4020)$, as well as their strange partners $Z_{cs}(3985)$ and $Z_{cs}(4120)$, on the basis of heavy quark spin symmetry. With an overall short-ranged contact potential in SU(3) flavor symmetry we show that the pole structures can be extracted from the experimental data. Meanwhile, we show the key role played by the $S$-wave open thresholds in $e^+e^-$ annihilations which are correlated with the production of these $Z$ states in a hadronic molecule picture. Implications of their scalar heavy quark spin partners $W_{c0,1,2}$ and $W_{cs0,1,2}$ are also gained.

\end{abstract}

\maketitle

\section{Introduction}

 Since the observations of $Z_c(3900)$~\cite{Ablikim:2013mio} and $Zc(4020)$~\cite{Ablikim:2013wzq} at BESIII the possible existence of their strange partners has been a focus of both experimental and theoretical studies (see e.g. several recent reviews on the relevant subject~\cite{Chen:2016qju,Esposito:2016noz,Guo:2017jvc,Ali:2017jda,Liu:2019zoy,Olsen:2017bmm}).
In Ref.~\cite{Ablikim:2020hsk} a new exotic candidate $Z_{cs}(3982)$ is reported by the BESIII Collaboration in $e^+e^-\to K^+ (D_s^-D^{*0}+D_s^{*-}D^0)$  in the invariant mass spectra of $D_s^-D^{*0}$ and $D_s^{*-}D^0$. It has a mass of $M_{Z_{cs}}=(3982.5^{+1.8}_{-2.6}\pm 2.1)$ MeV and a width of $(12.8^{+5.3}_{-4.4}\pm 3.0)$ MeV. Its mass is about 10 MeV  above the thresholds of $D_s^-D^{*0}$ and $D_s^{*-}D^0$ and makes it an ideal candidate for a hadronic molecule state of $D_s^-D^{*0}$ and $D_s^{*-}D^0$ as the strange partner of $Z_c(3900)$. This observation immediately initiates theoretical interests in uderstanding its nature~\cite{1830582,1830623,1830632,1830580,1830601,1830601}.

Theoretical prescriptions of the charged-charmonium states can be categorized into three groups by considering the analytical structures in terms of dynamical amplitudes in the complex energy plane: (i) 4-quark states in an overall color singlet; (ii) Hadronic molecules composed of two color-singlet hadrons; (iii) Kinematic effects which do not have pole structures~\footnote{Here, kinematic effects are referred to the leading triangle singularity mechanism. We do not consider the CUSP effects caused by an $S$-wave two-body branch points. The reason is that for weak couplings the two-body rescattering would not produce any narrow peak at the threshold. If the couplings become strong, it always requires a sum of infinite rescattering series to keep unitarity. This will then dynamically generate a pole and become dynamical~\cite{Guo:2014iya}. }. Although these three scenarios are motivated by different considerations of the underlying dynamics or mechanisms, they tackle different aspects of the non-perturbative QCD phenomena and in most cases it is hard to distinguish them in a specific case.

In association with the observations of $Z_c(3900)$, $Z_c(4020)$ and $Z_{cs}(3985)$ at BESIII~\cite{Ablikim:2013mio,Ablikim:2013wzq,Ablikim:2020hsk} and $Z_b^\pm(10610)$, $Z_b^\pm(10650)$ at Belle~\cite{Belle:2011aa} one observes a strong correlation between the open charm $S$-wave thresholds in $e^+e^-$ annihilations and the production mechanisms for these charged-charmonium states. In the charm sector it was first pointed out in Ref.~\cite{Wang:2013cya} that the production of $Z_c(3900)$ and $Z_c(4020)$ are correlated with the first narrow $S$-wave open channel $D_1(2420)\bar{D}+c.c.$ and $D_1(2420)\bar{D}^*+c.c.$, respectively, and can be related to each other by the heavy quark spin symmetry. The $S$-wave $D_1(2420)\bar{D}+c.c.$ and $D_1(2420)\bar{D}^*+c.c.$ couplings can explain the mysterious nature of $Y(4260)$ and $Y(4360)$ as hadronic molecules, thus, can accommodate these charmonium-like states under the same framework as hadronic molecules~\cite{Guo:2017jvc}. Furthermore, it was stressed in Ref.~\cite{Wang:2013cya} that the productions of the $Z_c$s and $Z_b$s are strongly enhanced by the triangle singularity (TS) mechanism which is a novel phenomenon in hadron transitions. This scenario was elaborated in a series of follow-up studies~\cite{Wang:2013hga,Cleven:2013mka,Qin:2016spb,Xue:2017xpu,Liu:2013vfa,Cao:2017lui} for the production of $Z_c(3900)$ and/or $Z_c(4020)$ and extended to the production of $Z_{cs}$~\cite{Cao:2017lui}. The observation of $Z_{cs}(3982)$ can be regarded as a manifestation of both the importance of the $S$-wave open threshold in $e^+e^-$ annihilations and the key role played by the TS mechanism. 

In this work we combine the production of $Z_c(3900)$ and $Z_c(4020)$ to demonstrate that the same mechanism gives rise to the $Z_{cs}$ signals in the elastic channels which suggests that there exist genuine pole structures for these states.   As follows, we first introduce the frameworks and then present the calculation results. A brief summary is given in the end. 


\section{Framework}

In hadronic molecular picture the property of the near-threshold states are largely correlated to the $S$-wave scattering of the relevant hadrons. For instance, the scattering between the $s_l^P=\frac{1}{2}^-$ doublet and its charged conjugate 
partners has been studied in both isospin singlet and triplet channel to explain the $X(3872)$~\cite{AlFiky:2005jd,Nieves:2012tt,Liu:2019stu,Nieves:2012tt,Baru:2016iwj,Baru:2015tfa},
$Z_c(3900)$ and $Z_c(4020)$~\cite{Aceti:2014uea,He:2017lhy,Xu:2017tsr,Sakai:2017avl,Pilloni:2016obd,Gong:2016hlt,Albaladejo:2015lob,He:2015mja,Zhao:2015mga,Zhao:2014gqa} in the hidden charm sector, and $Z_b(10610)$ and $Z_b(10650)$ \cite{Baru:2019xnh,Wang:2018jlv,Kang:2016ezb,Bicudo:2015kna,Dias:2014pva,Wang:2018atz,Dias:2014pva,Sun:2012zzda,Sun:2011uh,Nieves:2011zz} in the hidden bottom sector. Although the neutral $Z_c^{(\prime)0}$ and the $Z^{(\prime)}_{c\bar{c}s\bar{s}}$ can involve dynamics from the light meson exchanged potential due to the SU(3) breaking effect~\cite{Aceti:2014uea},
the charged states, $Z_c^{(\prime)\pm}$ and the $Z_{cs}^{(\prime)\pm}$ as well as their spin partners, only involve
short-ranged contact potentials, which are generally treated as input of an effective field theory (EFT)~\cite{Mehen:2011yh} based on their heavy quark spin structures~\cite{Bondar:2011ev}. In this study we adopt the same method~\cite{Hanhart:2015cua,Guo:2016bjq,Wang:2018jlv,Baru:2019xnh,Ohkoda:2012rj,Xiao:2013yca} for an overall description of the scatterings between the $s_l^P=\frac{1}{2}^-$ doublet for the $Z$ states.

The potential for the $Z$ states~\cite{Hanhart:2015cua,Guo:2016bjq,Wang:2018jlv,Baru:2019xnh,Voloshin:2011qa,Baru:2017gwo} can be expressed as
\begin{eqnarray}
V_{Z}&=&\frac{1}{2}\left(\begin{array}{cc}
\mathcal{C}_1+\mathcal{C}_0 & \mathcal{C}_1-\mathcal{C}_0\\
\mathcal{C}_1-\mathcal{C}_0& \mathcal{C}_1+\mathcal{C}_0
\end{array}\right)
\end{eqnarray}
where $C_0$ and $C_1$ are the two potential strengths. Similarly, the potential for the charge conjugation partners of $Z_c$s, i.e. $W_0$, $W_1$ and $W_2$, can be expressed in term of $C_0$ and $C_1$ in the heavy quark symmetry. 

As the potentials do not depend on energy and momentum, the Lippmann-Schwinger equation (LSE)
\begin{eqnarray}
\mathcal{F}_{\mathrm{phy}}=\mathcal{F}_{\mathrm{bare}}+\mathcal{F}_{\mathrm{phy}}G_\Lambda V,
\end{eqnarray}
with $\mathcal{F}_{\mathrm{bare}}$ and $\mathcal{F}_{\mathrm{phy}}$
the bare and physical production amplitudes, respectively, can be solved algebraically.
Here, the two-body non-relativistic propagator is
\begin{eqnarray*}
&&G_\Lambda(M)=\int\frac{\mathrm{d}^3q}{(2\pi)^3}\frac{1}{M-m_1-m_2-\vec{q}^2/(2\mu)}\\\nonumber
&&=\Lambda+i\frac{m_1 m_2}{2\pi(m_1+m_2)}\sqrt{2\mu (M-m_1-m_2)} \ ,
\end{eqnarray*}
where the power divergence subtraction~\cite{Kaplan:1998tg} is adopted to regularize the ultraviolet (UV) divergence;
$m_1$, $m_2$ and $\mu$ denote the masses of the intermediated two particles and their reduced mass, and $M$ is the  total energy of the system.

As mentioned earlier, the productions of the $Z_c$s and $Z_{cs}$s turn out to be strongly correlated with the $S$-wave open thresholds in $e^+e^-$ annihilations. In particular, a series of channels opened by the $\frac{1}{2}^+-\frac{1}{2}^-$ and $\frac{3}{2}^+-\frac{1}{2}^-$ pairs between $4.2\sim 4.7~\mathrm{GeV}$ indicate non-trivial mechanisms in association with the productions of these $Z$ states via the triangle scatterings. Although, dynamical calculations of the relevant open thresholds are needed for demonstrating the transition mechanisms, we adopt a parametrization scheme for a fixed center-of-mass (c.m.) energy in $e^+e^-$ annihilations.

Taking the production of $Z_c$s as an example we parametrize the bare production amplitudes of $D\bar{D}^*+c.c.$ and $D^*\bar{D}^*$ with the pion emission in a $D$ wave as 
\begin{eqnarray}\label{parametrize-D}
\mathcal{F}_{\pi_D}=\left(1,\mathcal{F}_D\right) \ ,
\end{eqnarray}
where the bare production amplitude of $D\bar{D}^*+c.c.$ is set to $1$ as a normalization. This contribution can be originated from the narrow $D_1(2420)\bar{D}+c.c.$ and $D_1(2420)\bar{D}^*+c.c.$ thresholds where the $D_1(2420)$ decays into $D^*\pi$ is via a $D$ wave~\cite{Wang:2013cya,Wang:2013hga,Cleven:2013mka,Qin:2016spb,Xue:2017xpu,Liu:2013vfa,Cao:2017lui}. 
Similarly, the bare production amplitudes of $D\bar{D}^*+c.c.$ and $D^*\bar{D}^*$ with the pion emission in an $S$ wave
can be parameterized as
\begin{eqnarray}\label{parametrize-S}
 \mathcal{F}_{\pi_S}=\left(\mathcal{F}_{S1},\mathcal{F}_{S2}\right) \ ,
\end{eqnarray}
where the $S$-wave contributions can come from the broad $D_1(2430)\bar{D}+c.c.$ and $D_1(2430)\bar{D}^*+c.c.$ thresholds with the $D_1(2430)$ dominantly decays into $D^*\pi$ via an $S$ wave~\cite{Cleven:2013mka,Qin:2016spb}. Also, the broad threshold $D_0(2400)\bar{D}^*$ can contribute to the final $D^*\bar{D}\pi+c.c.$ channel~\cite{Qin:2016spb}.
Then, the physical production amplitudes $\mathcal{F}_{\mathrm{phy}}^D$ and $\mathcal{F}_{\mathrm{phy}}^S$ can be obtained by solving the LSE. 

The physical production amplitudes $\mathcal{M}_{D\bar{D}^{*}\pi}$ and $\mathcal{M}_{D^{*}\bar{D}^{*}\pi}$ can be expressed as
\begin{widetext}
\begin{eqnarray}
\mathcal{M}_{D\bar{D}^{*}\pi}	&=&\epsilon_{Y}^{a}\epsilon_{\bar{D}^{*}}^{*b}\left(\mathcal{F}_{\mathrm{phy}}^{S1}\delta^{ab}+\mathcal{F}_{\mathrm{phy}}^{D1}\left(\hat{p}_{\pi}^{*a}\hat{p}_{\pi}^{*b}-\frac{1}{3}\delta^{ab}\right)p_{\pi}^{*2}\right),\\
\mathcal{M}_{D^{*}\bar{D}^{*}\pi}	&=&\epsilon_{Y}^{a}\frac{i}{\sqrt{2}}\epsilon^{bcd}\epsilon_{D^{*}}^{*c}\epsilon_{\bar{D}^{*}}^{*d}\left(\mathcal{F}_{\mathrm{phy}}^{S2}\delta^{ab}+\mathcal{F}_{\mathrm{phy}}^{D2}\left(\hat{p}_{\pi}^{*a}\hat{p}_{\pi}^{*b}-\frac{1}{3}\delta^{ab}\right)p_{\pi}^{*2}\right),
\end{eqnarray}
\end{widetext}
where $\epsilon_Y$ is the polarization of the intermediate vector state produced in $e^+e^-$ annihilations, and it satisfies
\begin{eqnarray}
\sum_{\lambda=1,2}\epsilon_{Y}^{\lambda a}\epsilon_{Y}^{*\lambda b}=\delta^{ab}-\delta^{a3}\delta^{b3} \ ,
\end{eqnarray}
with the third polarization component absent. 
Accordingly, the cross section will be proportional to 
\begin{eqnarray}\nonumber
\sum_{\text{polarizations}}|\mathcal{M}_{D\bar{D}^{*}\pi}|^{2}	=2|\mathcal{F}_{\text{phy}}^{S1}|^{2}+2\text{Re}[\mathcal{F}_{\text{phy}}^{S1}\mathcal{F}_{\text{phy}}^{D1*}]\left(\frac{1}{3}-\cos^{2}(\theta)\right)p_{\pi}^{2}+|\mathcal{F}_{\text{phy}}^{D1}|^{2}\left(\frac{5}{9}-\frac{1}{3}\cos^{2}(\theta)\right)p_{\pi}^{4},
\end{eqnarray}
where $p_\pi^*$ is the three momentum of the emitted pion in the c.m. frame and $\theta$ is the angle between the spectator pion and the beam axis in the same frame.  
The $D^{*}\bar{D}^{*}\pi$ channel can be obtained by replacing $\mathcal{F}_{\text{phy}}^{S1}$, $\mathcal{F}_{\text{phy}}^{D1}$ by $\mathcal{F}_{\text{phy}}^{S2}$, $\mathcal{F}_{\text{phy}}^{D2}$, respectively. 

Thus, the differential partial widths are
\begin{eqnarray}
\frac{d\Gamma(D^{(*)}\bar{D}^*\pi)}{dMd\cos(\theta)}&=&\frac{1}{3}\frac{2m_{D^{(*)}}2m_{\bar{D}^{*}}2E}{32\pi^{3}E^{2}}p_{\pi}^{*}k|\mathcal{M}_{D^{(*)}\bar{D}^*\pi}|^{2},\label{eq:Swave}
\end{eqnarray}
where $E=\sqrt{s}$ is the overall c.m. energy, and $k$ is the three momentum of the $D^{(*)}$ in the $D^{(*)}\bar{D}^{(*)}$ rest frame.

With the partial width distributions above, we perform three fitting schemes: 
\begin{itemize}
\item Scheme I: only the $S$-wave pion emmission considered;
\item Scheme II: only the $D$-wave pion emission considered;
\item Scheme III: both the $S$ and $D$-wave pion emissions considered which is the sum of these two partial waves.
\end{itemize}
The idea is that for a fixed energy, in particular, with the masses close to the initial vector charmonium or charmonium-like states associated with the nearby open threshold, the production mechanisms will be encoded in the parameters defined in Eqs.~(\ref{parametrize-D}) and (\ref{parametrize-S}), and the pole structure of the corresponding $Z$ state can be extracted by the lineshape of the partial width distributions. 



\begin{figure*}[!htb]
 \centering
  \includegraphics[width=0.9\textwidth]{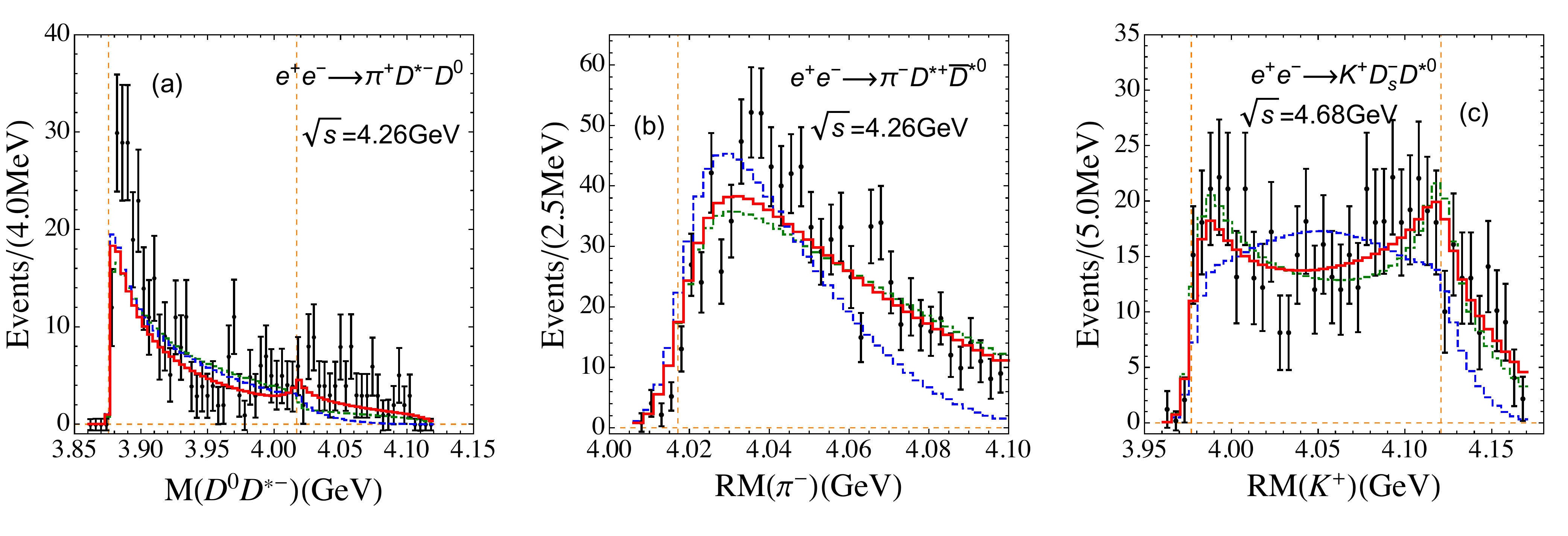}
  \caption{The fitted invariant mass distributions of the interested system
   in the $e^+e^-\to \pi^+ D^{*-}D^0$, $e^+e^-\to\pi^- D^{*+}\bar{D}^{*0}$,
   $e^+e^-\to K^+ D_s^-D^{*0}$ processes. 
  The green dot-dashed, blue dashed and red solid curves are the 
  fitted results for Schemes I, II and III  with $\chi^2/\mathrm{d.o.f}=1.4, 2.3, 1.2$, respectively.
    The vertical orange dashed lines in (a), (b) are the $D\bar{D}^*+c.c.$
  and $D^*\bar{D}^*$ thresholds. Those in (c)
    are the $D_s^-D^{*0}$ and $D_s^{*-}D^{*0}$ thresholds. 
  The corresponding experimental data are taken from 
  Refs.~\cite{Ablikim:2015swa,Ablikim:2013emm,Ablikim:2020hsk}, 
  respectively. The resolutions have been considered for each channel.
        }
  \label{fig1}
\end{figure*}

\begin{figure*}[!htb]
 \centering
  \includegraphics[width=0.9\textwidth]{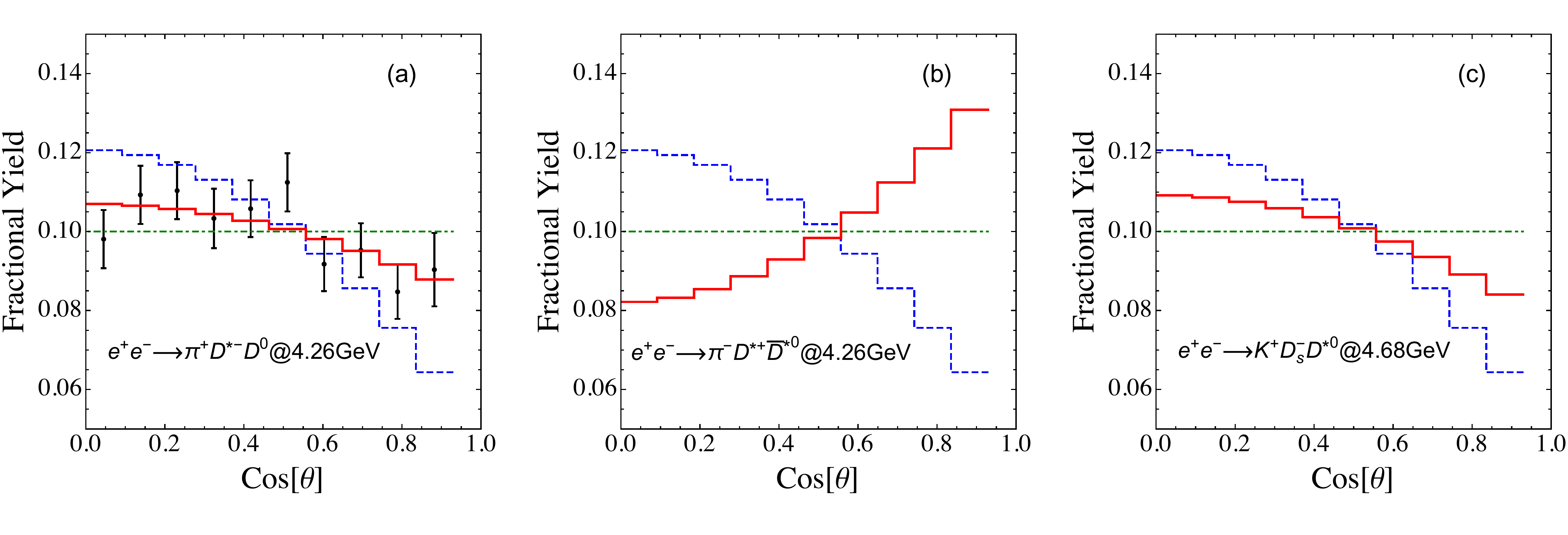}
  \caption{The Jackon angle distribution of
   the $e^+e^-\to \pi^+ D^{*-}D^0$, $e^+e^-\to\pi^- D^{*+}\bar{D}^{*0}$,
   $e^+e^-\to K^+ D_s^-D^{*0}$ processes at the first peak position,
   i.e. the mass region of $Z_c$, $Z_c^\prime$ and $Z_{cs}$ respectively, are presented.
  The green dot-dashed, blue dashed and red solid curves are the 
  fitted results for Schemes I, II and III. 
  The experimental data in (a) are taken from 
  Refs.~\cite{Ablikim:2015swa} and included in the fit.
  The angular distribution in the last two processes
  are presented as the prediction which can be used for the comparison with further experimental measurement.
        }
  \label{fig2}
\end{figure*}

\begin{table*}
\caption{The poles on the physical sheets and those close to the physical ones are listed. 
Those indicated by ``$--$" are either far away from the corresponding threshold (more than $100~\mathrm{MeV}$)
or on a sheet far away from the physical one. 
The subscripts ``B" and ``V" denote bound state on the physical sheet and virtual state on the sheet close to the physical one, respectively.}
\begin{tabular}{|c|c|c|c|c|c|c|c|}
\hline 
\multirow{2}{*}{States} & \multicolumn{3}{c|}{Poles} & \multirow{2}{*}{States} & \multicolumn{3}{c|}{Poles}\tabularnewline
\cline{2-4} \cline{3-4} \cline{4-4} \cline{6-8} \cline{7-8} \cline{8-8} 
 & Scheme I & Scheme II & Scheme III &  & Scheme I & Scheme II & Scheme III\tabularnewline
\hline 
\hline 
$Z_{c}$ & $3873.11_{\mathrm{V}}$ & $3875.76_{\mathrm{V}}$ & $3800.58_{\mathrm{B}}$ & $Z_{cs}$ & $3976.68_{\mathrm{V}}$ & $3979.39_{\mathrm{V}}$ & $3916.19_{\mathrm{B}}$\tabularnewline
\hline 
$Z_{c}^{\prime}$ & $--$ & $--$ & $--$ & $Z_{cs}^{\prime}$ & $--$ & $--$ & $--$\tabularnewline
\hline 
$W_{c0}$ & $3734.17_{\mathrm{B}}$ & $3702.63_{\mathrm{B}}$ & $3687.42_{\mathrm{B}}$ & $W_{cs0}$ & $3835.36_{\mathrm{B}}$ & $3806.73_{\mathrm{B}}$ & $3805.43_{\mathrm{B}}$\tabularnewline
\hline 
$W_{c0}^{\prime}$ & $--$ & $4022.07\pm i6.58$ & $--$ & $W_{cs0}^{\prime}$ & $--$ & $4125.26\pm i5.55$ & $--$\tabularnewline
\hline 
$W_{c1}$ & $3869.09_{\mathrm{B}}$ & $--$ & $--$ & $W_{cs1}$ & $3970.74_{\mathrm{B}}$ & $--$ & $--$\tabularnewline
\hline 
$W_{c2}$ & $4011.14_{\mathrm{B}}$ & $--$ & $--$ & $W_{cs2}$ & $4115.21_{\mathrm{B}}$ & $--$ & $--$\tabularnewline
\hline 
\end{tabular}
\label{tab1}
\end{table*}

\section{Results and Discussions}
The fitted invariant mass distributions and Jackon angular distributions for these three fitting schemes
are presented by the green dot-dashed, blue dashed, and red curves, respectively, 
in Figs.~\ref{fig1} and ~\ref{fig2}. 
The fit of Scheme III is much better than 
those of the others.
There are significant narrow peak structures above the $D\bar{D}^*$ and $D_s\bar{D}^*$ thresholds 
in the elastic channels, indicating poles~\cite{Guo:2014iya} as listed in Table~\ref{tab1}. 
The lineshapes around the $D^*\bar{D}^*$ and $D^*_s\bar{D}^*$ thresholds
show just a normal cusp effect due to the corresponding poles either with large distance to
the corresponding thresholds or on the sheets far away from the physical ones. 
Those are the reasons why they do not show significant impact on the physical 
observables.

With the parameters extracted from fitting to the invariant mass distributions and the Jackon anglular distributions,
one can extract the pole positions as shown in Table~\ref{tab1}.
The $Z_c(3900)$ in the first two fitting schemes
are virtual states with distances of $2.7~\mathrm{MeV}$ and $5~\mathrm{keV}$ to the $D\bar{D}^*$ threshold, respectively.
The accumulation of the events at the lower $D\bar{D}^*$ invariant mass 
for the $D$-wave makes the pole not necessary very close to the threshold. 
That is the reason why the binding energy of the $Z_c(3900)$ in Scheme III turns out to be large, i.e. $75.2~\mathrm{MeV}$. 
The poles of the $Z_c(4020)$ in the first two schemes are on the sheet far away from the physical one. 
In contrast, it appears as a deep virtual state in scheme III, which is beyond the acceptance of the molecular picture. 
It is interesting to note that these three fitting schemes have demonstrated that $Z_c(3900)$ exists as a genuine state, which is consistent with the conclusions of Refs.~\cite{Cleven:2013mka,Qin:2016spb}. Meanwhile, the fitting indicates $Z_{cs}(3985)$ exists as a genuine state which can be either a virtual state or a bound state. In Fig.~\ref{fig2} we make a prediction of the missing kaon spectrum for the potential $Z_{cs}(4120)$ as a hadronic molecule of $D_s^{*-}D^{*0}$ in these three schemes.

\begin{figure}[!htb]
 \centering
  \includegraphics[width=0.5\textwidth]{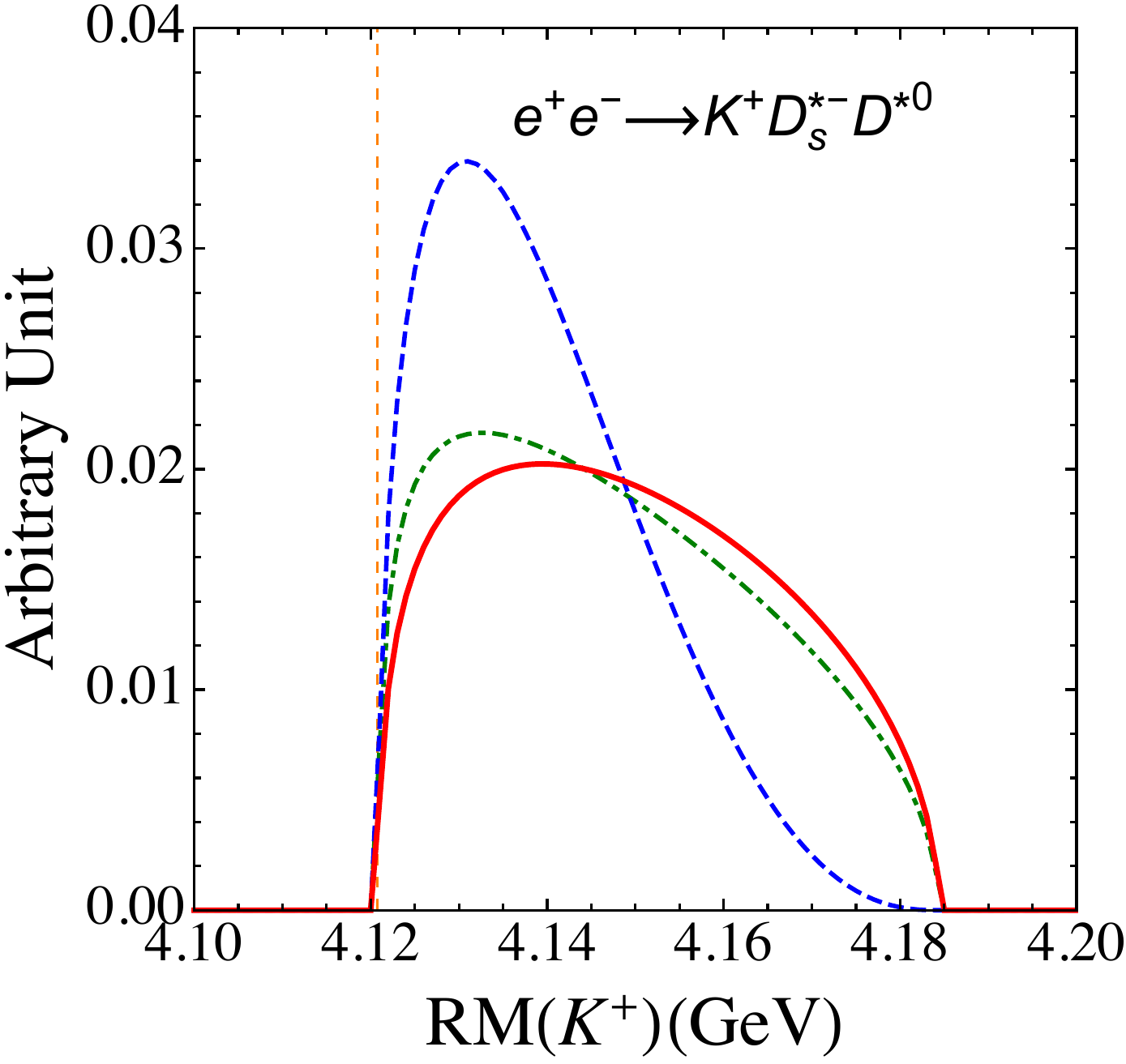}
  \caption{The predicted missing kaon distribution
  of the $e^+e^-\to K^+D_s^{*-}D^{*0}$ process at
  the c.m. energy $4.68~\mathrm{GeV}$.
  The blue dashed, green dot-dashed, and red curves are for Schemes I, II and III, respectively. }
  \label{fig3}
\end{figure}

As a by-product, we find that the HQSS partner of the $Z_c$ and $Z_{cs}$ states, $W_{c0}$ ($W_{cs0}$), behave as bound states on the physical sheet with binding energies 
$0.3~\mathrm{MeV}$, $31.8~\mathrm{MeV}$ and $47.0~\mathrm{MeV}$ ($0.001~\mathrm{MeV}$, $28.6~\mathrm{MeV}$ and $29.9~\mathrm{MeV}$), respectively. 
Due to its heavy quark spin structure, it can be searched in the decay channels of 
$J/\psi\pi\pi$ and $\eta_c\pi$. The $W_{c0}^\prime$ only exists in scheme II as a resonance.
We find that $W_{c1}$ and the $W_{c2}$ only exist as bound states, in scheme I. 
The presence of various $S$-wave thresholds 
in $e^+e^-$ between $4.0\sim 5.0~\mathrm{GeV}$ 
increases the difficulty of uncovering the nature of the $Z_c^{(\prime)}$ and  $Z_{cs}^{(\prime)}$,
comparing to their bottomonium analogies. 

\section{Summary and Outlook}
In this work we perform a coherent study of the production of these newly observed charged-charmonium $Z_c$ and $Z_{cs}$ states in $e^+e^-$ annihilations. While we emphasizing the key role played by the $S$-wave open thresholds in the understanding of their production mechanism, we parametrize out the mechanisms by the $S$ and $D$-wave production amplitudes, which are largely constrained by the Jackon angular distributions,
 in terms of the recoiled pion or kaon in the final states.  Then the lineshapes of the invariant mass spectra can provide a constraint on the pole structures of these exotic candidates. We find that $Z_c(3900)$ and $Z_{cs}(3982)$ both exist as genuine states, either virtual states or bound states. In addition, we find the spin partners of the $Z_c$ ($Z_{cs}$) states, $W_0$ ($W_{cs0}$), can exist as a bound state which can be searched for in $J/\psi\pi\pi$ ($J/\psi K\pi$) and $\eta_c\pi$ ($\eta_c K$) channels.

\medskip

\begin{acknowledgements}
This work is supported, in part, by the National Natural Science Foundation of China (Grant Nos. 11425525, 11521505 and 12035007), DFG and NSFC funds to the Sino-German CRC 110 ``Symmetries and the Emergence of Structure in QCD'' (NSFC Grant No. 11621131001 and DFG Grant No. TRR110), Strategic Priority Research Program of Chinese Academy of Sciences (Grant No. XDB34030302).
Q.W. is also supported by Science and Technology Program of Guangzhou (No. 2019050001) and Guangdong Provincial funding with No.2019QN01X172.

\end{acknowledgements}

\end{document}